\documentclass[aps,prl,preprint,superscriptaddress,showpacs,floatfix]{revtex4-1}

\usepackage{times}
\usepackage{graphicx}
\usepackage{dcolumn}
\usepackage{bm}
\usepackage{color}
\usepackage{longtable}
\usepackage{booktabs}
\usepackage{amssymb}
\usepackage{textcomp}

\begin{document}

\title{Thermodynamics around the First-Order Ferromagnetic Phase Transition of $\mathbf{Fe_{2}P}$ Single Crystals}

\author{M. Hudl}\email{matthias.hudl@mat.ethz.ch, hudl@kth.se}
\affiliation{Department of Materials, ETH Z\"{u}rich, Vladimir-Prelog-Weg 4, CH-8093 Zurich, Switzerland}
\affiliation{KTH Royal Institute of Technology, ICT Materials Physics, Electrum 229, SE-164 40 Kista, Sweden}

\author{D. Campanini}
\affiliation{Department of Physics, Stockholm University, SE-106 91 Stockholm, Sweden}

\author{L. Caron}
\affiliation{Fundamental Aspects of Materials and Energy, Faculty of Applied Sciences, TU Delft Mekelweg 15, 2629 JB Delft, The Netherlands}

\author{V. H\"{o}glin}
\affiliation{Department of Chemistry - \AA ngstr\"{o}m, Uppsala University, Box 538, SE-751 21 Uppsala, Sweden}

\author{M. Sahlberg}
\affiliation{Department of Chemistry - \AA ngstr\"{o}m, Uppsala University, Box 538, SE-751 21 Uppsala, Sweden}

\author{P. Nordblad}
\affiliation{Department of Engineering Sciences, Uppsala University, Box 534, SE-751 21 Uppsala, Sweden}

\author{A. Rydh}\email{andreas.rydh@fysik.su.se}
\affiliation{Department of Physics, Stockholm University, SE-106 91 Stockholm, Sweden}

\date{\today}

\begin{abstract}
The specific heat and thermodynamics of Fe$_2$P single-crystals around the first order paramagnetic (PM) to ferromagnetic (FM) phase transition at $T_{\mathrm C}\,\simeq\,217\,{\mathrm K}$ are empirically investigated. The magnitude and direction of the magnetic field relative to the crystal axes govern the derived $H$-$T$ phase diagram. Strikingly different phase contours are obtained for fields applied parallel and perpendicular to the $c$-axis of the crystal. In parallel fields, the FM state is stabilized, while in perpendicular fields, the phase transition is split into two, with an intermediate FM phase where there is no spontaneous magnetization along the $c$-axis. The zero-field transition displays a text-book example of a first order transition with different phase stability limits on heating and cooling. The results have special significance since Fe$_2$P is the parent material to a family of compounds with outstanding magnetocaloric properties. 

\end{abstract}


\maketitle

First-order magnetic phase transitions (FOMT) attract attention due to their promising applications in magnetic shape memory alloys \cite{Chernenko2008, Enkovaara2004}, magnetic sensing based on colossal magnetoresistance \cite{Ramirez1997}, and magnetic refrigeration \cite{Tishin03,Gschneidner05,Shen09,deOliveira10,Wang:2012wg}.
The FOMT may be driven by temperature, pressure or applied magnetic field, and is associated with a sudden change of structure or lattice parameters, resistivity, and magnetic entropy. Anisotropy and structural distortions play important roles for the FOMT. Materials with strong magnetic anisotropy, for instance, enable unusual and highly efficient ways in tuning magnetocaloric effects \cite{Reis08,Nikitin10}. Di-iron phosphide, Fe$_2$P, is a hexagonal transition metal pnictide with strong magneto-crystalline anisotropy that undergoes a first-order paramagnetic to ferromagnetic phase transition at $\sim216\,{\mathrm K}$ \cite{Wappling75,Fujii77,Lundgren78}. The transition of Fe$_2$P is magneto-elastic, with a discontinuous iso-structural change of the dimensions of the hexagonal unit cell \cite{Lundgren78} at the Curie temperature $T_{\mathrm C}$. Below $T_{\mathrm C}$ the magnetic ordering occurs along the crystallograhic $c$-axis direction \cite{Fujii77}. The thermodynamics at the phase transition is of particular interest since Fe$_2$P is the parent compound for an entire class of tunable first-order magnetocaloric materials built on rare-earth-free constituents~\cite{Dung11}.

The anisotropic magnetic and magnetocaloric properties of Fe$_2$P were recently studied by Caron \emph{et\,al}.~\cite{Caron13b}. Specific heat for polycrystalline Fe$_2$P in zero magnetic field has been studied earlier~\cite{Beckman82}. A deeper thermodynamic understanding of the FOMT is, however, lacking. In this work we report high-resolution nanocalorimetric specific heat measurements of high-quality Fe$_2$P single crystals in applied magnetic fields both parallel and perpendicular to the $c$-axis. We furthermore directly probe the latent heat of the FOMT. The results provide new insights into the $H$-$T$ phase diagram of Fe$_2$P in the region around the magnetic phase transition, illuminating the behavior of first-order systems with strong magnetic anisotropy and present a textbook example of diverging specific heat around \emph{first order} transitions in general.

\begin{figure*}[ht!]
	\includegraphics[width=0.94\linewidth]{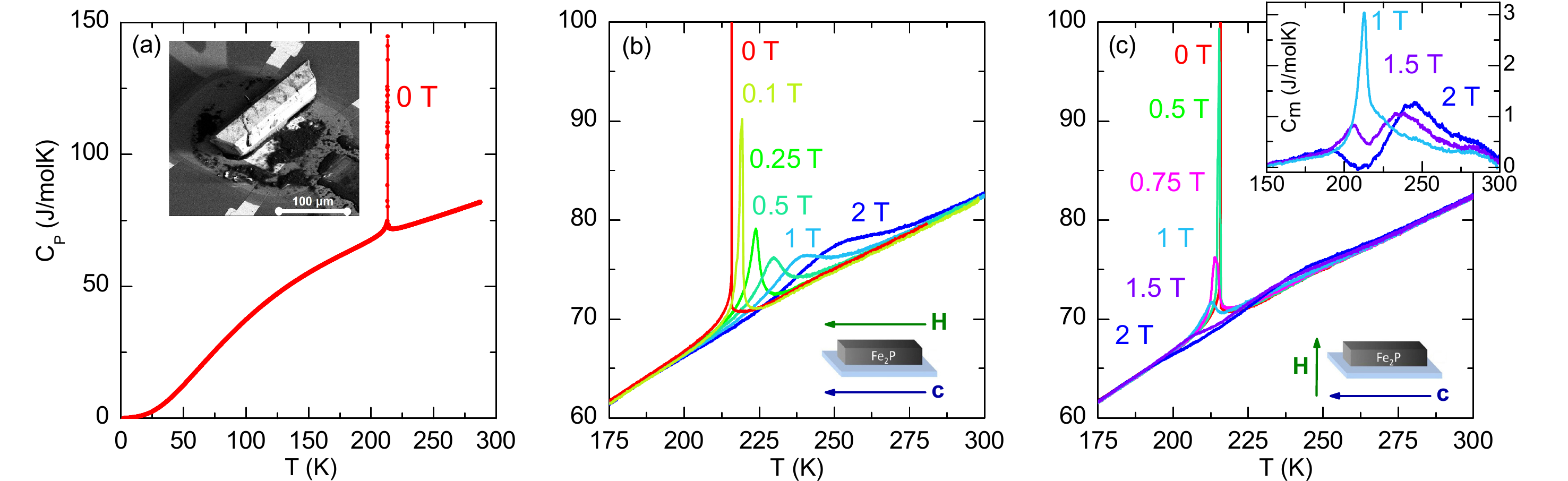}
	\caption{Temperature dependence of the specific heat of an Fe$_2$P crystal. (a) Specific heat in zero field. The inset shows the sample mounted on the nanocalorimeter membrane cell. (b) Specific heat with magnetic field along the $c$-axis (0, 0.1, 0.25, 0.5, 1 and 2 T).
(c) Specific heat with field perpendicular to the $c$-axis (0, 0.5, 0.75, 1, 1.5 and 2 T). The inset shows the magnetic contribution C$_\mathrm{m}$ at high fields (1, 1.5, and 2 T).}
	\label{Fe2P_Calori}
\end{figure*}

High-quality single crystals of Fe$_2$P were grown in a tin melt \cite{Madar77,Andersson78}. Single crystal x-ray diffraction intensities in the ferromagnetic phase at $100\,\mathrm{K}$ were recorded on a Bruker diffractometer. The composition was refined to be Fe$_{1.995(2)}$P, where only the Fe(2) site is fully occupied. Magnetization measurements were performed using a Quantum Design MPMS XL SQUID and PPMS $9\,\mathrm{T}$ VSM. Specific heat measurements were performed using a differential membrane-based nanocalorimeter applying an AC-method with phase stabilized frequency feedback \cite{Tagliati12}. The device allows measurements of \textmu g size samples with both high resolution ($\mathrm{\Delta C/C \leq 10^{-4}}$) and good absolute accuracy. Latent heat was measured with the same nanocalorimeter operating in a differential scanning mode. The single-crystal sample used for the specific heat measurements was pillar shaped (length $189\pm 9\,$\textmu m) with hexagonal basal plane (edges $31.9 \pm 4.3\,$\textmu m).

The temperature dependence of the zero-field specific heat $C_{\mathrm{p}}$ is shown in Fig.~\ref{Fe2P_Calori}(a). Specific heat values obtained by Beckman \emph{et\,al}.~\cite{Beckman82} at $150\,\mathrm{K}$ were used as a scale reference. The obtained sample mass agrees with the estimate from microscopy within uncertainties. The general temperature dependence and low-temperature behavior of the specific heat is in good agreement with the results in literature~\cite{Beckman82}. The specific heat associated with the FOMT appears as a sharp peak at the ferromagnetic-to-paramagnetic  transition with $T_\mathrm{C}\simeq 217\,\mathrm{K}$. The magnetic field dependence of the specific heat in the vicinity of the phase transition is shown in Fig.~\ref{Fe2P_Calori}(b,c). For fields $\bf{H}\, \parallel \, \bf{c}$, the transition is broadened and shifted to higher temperatures. When $\bf{H}\, \perp\, \bf{c}$, the location of the transition first appears nearly unchanged for fields below $0.5\,{\mathrm T}$ but is then split into two at higher fields, see Fig.~\ref{Fe2P_Calori}(c). The lower transition shifts down in temperature while the upper shifts to higher temperatures with increasing applied field, as is evident from the inset of Fig.~\ref{Fe2P_Calori}(c), showing the magnetic contribution $C_\mathrm{m}=C_\mathrm{tot}-C_\mathrm{bg}$ for high fields. The background specific heat $C_\mathrm{bg}$ is determined as an interpolation of the lowest specific heat for any field or field direction.

\begin{figure}[b]
	\includegraphics[width=0.594\linewidth]{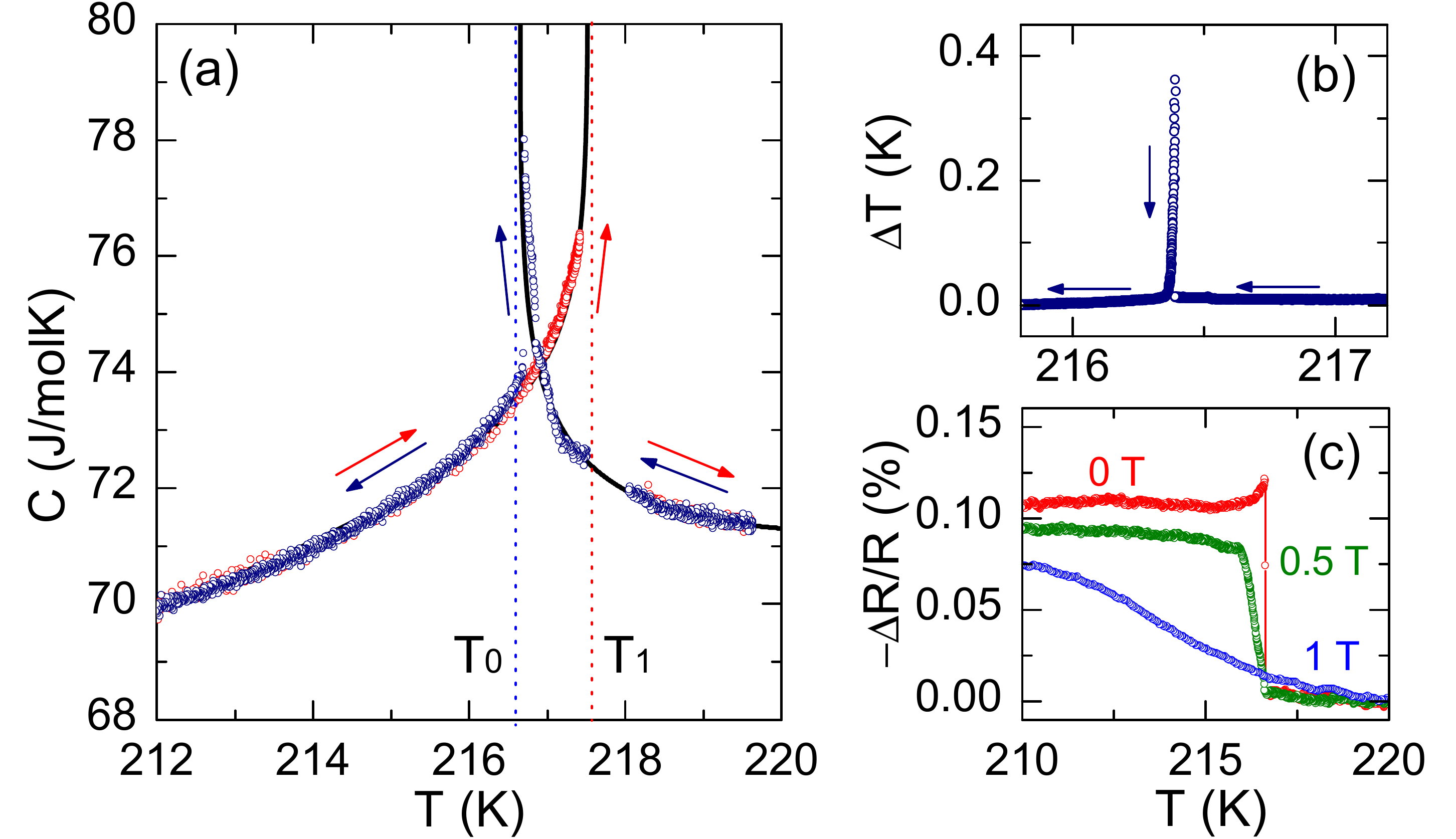}
	\caption{(a) Specific heat of an Fe$_2$P single-crystal around the zero-field transition. The paramagnetic state supercools to a temperature $T_0$ while the ferromagnetic state superheats to $T_1$. The specific heat diverges at these corresponding temperatures rather than at $T_\mathrm{C}$, following $C=C_0+C_1\cdot T+C_{\rm m}$ as shown by the solid curves. (b) Differential temperature $\Delta T = T_{\mathrm{sample}} - T_{\mathrm{ref}}$ as a function of reference temperature on cooling in zero field. A sudden temperature change of the sample is seen when the latent heat is released during the transition. (c) Persistent thermometer resistance change of the calorimeter due to the thermal expansion at the magnetoelastic transition of the sample for fields $\bf{H}\, \perp\, \bf{c}$.}
	\label{Fe2P_Latent}
\end{figure}

To investigate the character and latent heat of the FOMT, the specific heat in zero field was measured with a small temperature oscillation amplitude $T_\mathrm{ac}< 160\,\mathrm{mK}$ on very slow heating and cooling. As seen in Fig.~\ref{Fe2P_Latent}(a), there is a clear hysteresis between heating and cooling, indicating a first order transition \cite{Binder87,Roy2013}. By using the ac method with a small temperature oscillation amplitude on a small single crystal, the true specific heat is measured, without admixture of latent heat in the signal.  At the supercooling and superheating stability limits $T_0$ and $T_1$ the system undergoes an instantaneous transition between the involved states. To quantify the latent heat of the transition, the calorimeter was turned into differential scanning mode, where the temperature difference between sample and reference calorimetric cells were measured at high speed during a slow base temperature scan. The latent heat at the transition gives rise to a sudden adiabatic temperature change $\Delta T$ of the sample and a subsequent temperature relaxation, as shown in Fig.~\ref{Fe2P_Latent}(b). We find that the temperature increases by $\Delta T = 372\,\mathrm{mK}$ when going from high to low temperature in zero applied field, corresponding to a latent heat of $L=26.4\,{\mathrm{J/mol}}$ or an entropy change $\Delta S_{\mathrm{L}}\,=\,L/T_{\mathrm C}\,=\,122\,{\mathrm{mJ/mol\,K}}\approx 0.015\,\mathrm{R}$. In magnetic fields $\mathbf{H}\perp\,\bf{c}$ the latent heat first decreases and finally vanishes for $H$ slightly higher than $0.5\,{\mathrm T}$. At $0.5\,{\mathrm T}$, the latent heat is about 15\% of that in zero field. Magnetic fields $\mathbf{H}\parallel\,\bf{c}$, on the other hand, quickly suppress the first order character already below $0.1\,\mathrm{T}$.
As seen in Fig.~\ref{Fe2P_Latent}(a), there is a divergent behavior of the zero-field specific heat on
approaching the FOMT. However, this divergence does not occur around the equilibrium temperature $T_\mathrm{C}$, but at the stability limits $T_0$ and $T_1$. This is expected for a first order transition \cite{Binder87}. The temperature dependence of $C_\mathrm{m}$ is well described by a power law of the type  $C_{\mathrm{m}}= ({\Gamma_{\pm}}/{n}) \cdot{\left|\varepsilon\right|}^{-n}\label{Eq:Cm}$ \cite{Kim2002,KornblitAAhlers1973}, where $\varepsilon = 1 - T/T_i$, $n$ is an exponent, $i=0, 1$, and $\Gamma_{\pm}$ are fitting parameters for $\varepsilon >0$ (+) using $T_0=216.64\,\mathrm{K}$ and $\varepsilon <0$ (-) using $T_1=217.52\,\mathrm{K}$, respectively. 
From fitting, using data in the reduced temperature range $5\times 10^{-4} < \varepsilon < 10^{-2}$,  we find $n\, =\, -0.006\, \pm\, 0.01$ and a ratio $\Gamma_+/\Gamma_- \,=\, 1.006\, \pm\, 0.01$ (not to be confused with the critical values of second order phase transitions). 

When decreasing the temperature across the transition in zero magnetic field, the $c$-axis undergoes a sudden contraction by 0.084\% while the in-plane lattice parameter expands by 0.074\% \cite{Fujii77}. These thermal expansions can be sensed by the thin-film thermometer of the calorimeter, acting as a thermal expansion gauge. While this method does not give absolute values of volume changes, it provides a good account for the location and character of the iso-structural transition. Figure~\ref{Fe2P_Latent}(c) shows the signal from thermal expansion during cooling in magnetic fields $\bf{H}\, \perp\, \bf{c}$. As seen in the figure, the magnitude of the structural transition is suppressed by the magnetic field, and the transition is broadened and shifted towards lower temperatures. The structural change in a field of $0.5\,\mathrm{T}$ still amounts to more than 60\% of that in zero field, while the latent heat has been significantly decreased. From this, we conclude that the latent heat associated with the structural transition itself is at most a minor part ($\lesssim 10\%$) of the total latent heat.  

\begin{figure}[t]
	\includegraphics[width=0.594\linewidth]{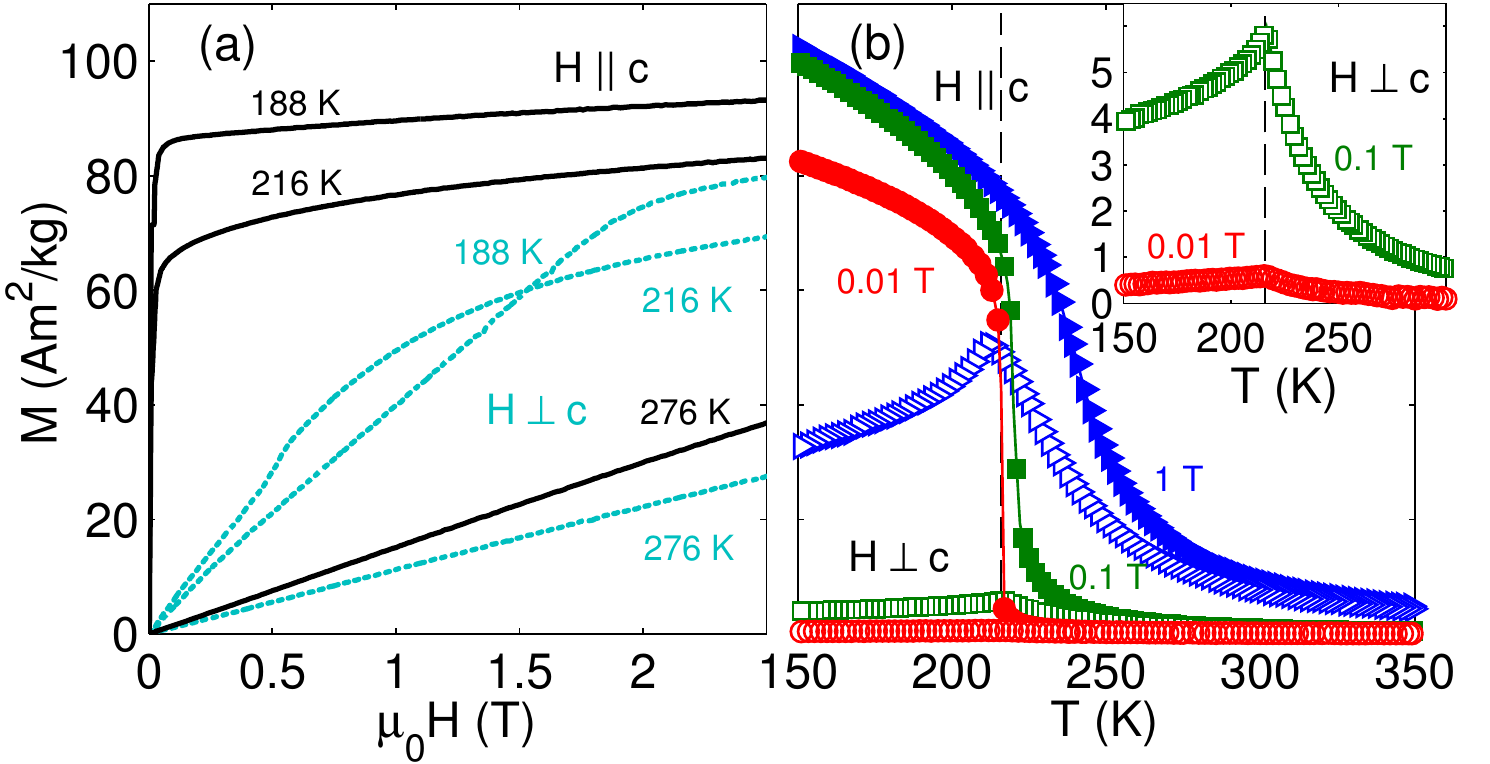}
	\caption{Magnetization along the applied field of an Fe$_2$P crystal. (a) Magnetization as function of field applied along the $c$-axis (solid curves) and perpendicular to the $c$-axis (dashed curves) for three different temperatures; $188\,\mathrm{K}$, $216\,\mathrm{K} \approx T_{\mathrm C}$ and $276\,\mathrm{K}$. (b) Temperature dependence of the magnetization for fields along (filled symbols) and perpendicular to (open symbols) the $c$-axis (\textcolor{red}{$\circ$} = 0.01 T,  \textcolor{green}{$\Box$} = 0.1 T, and \textcolor{blue}{$\rhd$} = 1 T). The inset shows the low-field curves for $\bf{H}\, \perp\, \bf{c}$. $T_\mathrm{C} (0\,{\rm T})$ is indicated by vertical dashed lines.}
	\label{Fe2P_Magnetization}
\end{figure}

To understand the nature of the observed transitions, the specific heat measurements were complemented with magnetization measurements. The field dependence of the magnetization is shown in Fig.~\ref{Fe2P_Magnetization}(a) for $\mathbf{H}\parallel\,\bf{c}$ (solid curves) and $\mathbf{H}\perp\,\bf{c}$ (dashed curves), and the corresponding temperature dependence is shown in Fig.~\ref{Fe2P_Magnetization}(b). The results agree well with Ref.~\cite{Caron13b}.
As seen from Fig.~\ref{Fe2P_Magnetization}, the system orders spontaneously along the $c$-axis below the Curie temperature $T_\mathrm{C}=217\,\mathrm{K}$. The sharp increase of magnetization at $T_\mathrm{C}$ is accompanied by thermal hysteresis ($\lesssim$ 1 K), indicating first-order character of the phase transition. With increasing fields along the $c$-axis, the transition temperature to the ferromagnetic state is quickly shifting to higher temperatures. However, for $\bf{H}\, \perp \, \bf{c}$ the magnetization suddenly starts to decrease below $T_\mathrm{C}$, creating a cusp-shaped magnetization curve. Comparable magnetic behavior are seen, e.g., for the para-to-ferromagnetic transitions in MnP \cite{Reis08} and DyAl$_2$ \cite{vonRanke00,Lima05,vonRanke07}. This behavior can be understood by considering the magneto-crystalline anisotropy. At low fields along the $c$-axis, the system orders spontaneously at $T_\mathrm{C}$. A perpendicular applied field ($\bf{H}\, \perp \, \bf{c}$) tilts the magnetization vector away from the $c$-axis. When the field reaches the anisotropy field, the $c$-axis component goes to zero. At constant field the magnetization $\bf{M}_{\perp \, \bf{c}}$ is suppressed below a certain temperature (and exhibits a cusp) due to an increasing magneto-crystalline anisotropy with decreasing temperature and the onset of magnetic ordering along the $c$-axis. 

\begin{figure}[b!]
	\includegraphics[width=0.594\linewidth]{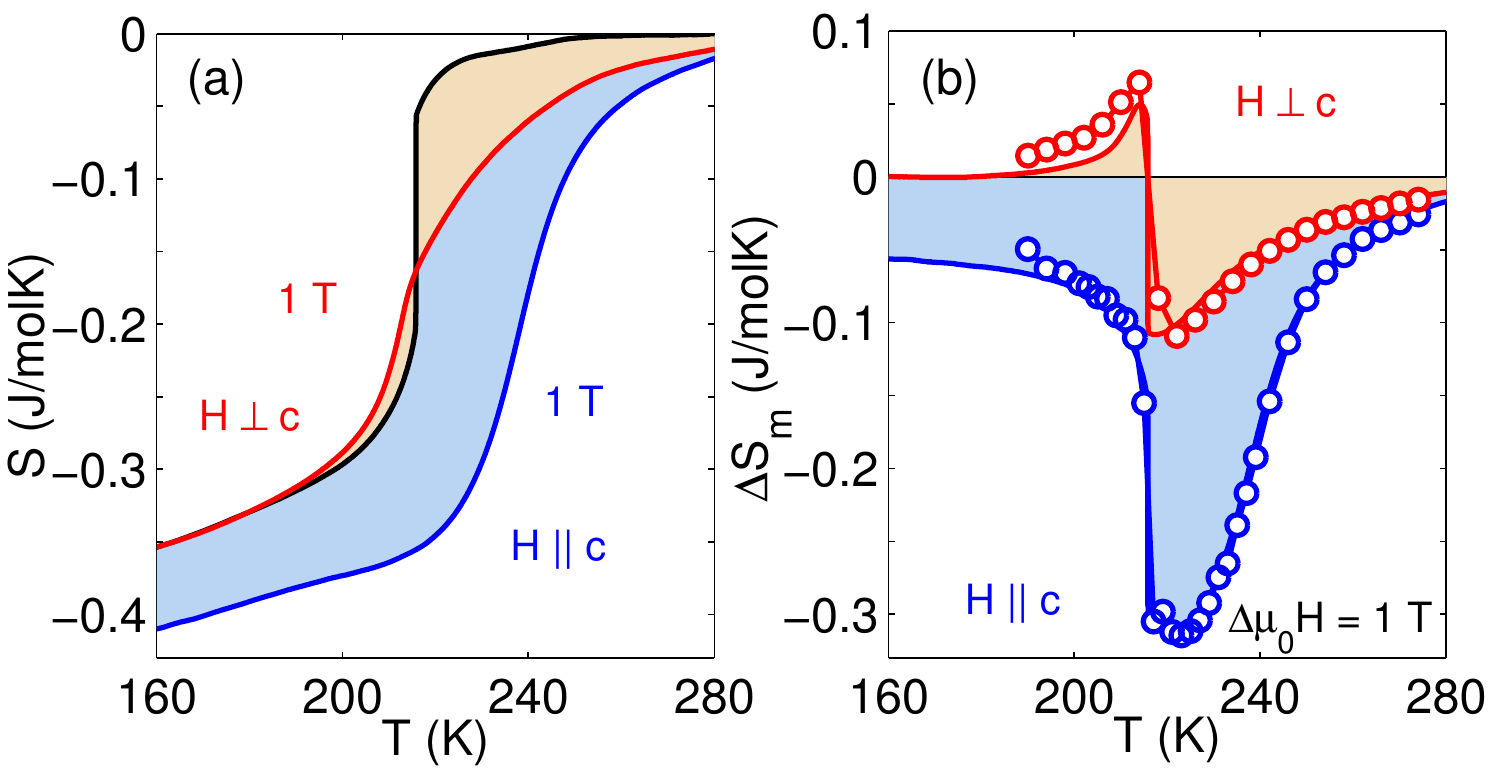}
	\caption{(a) Magnetic entropy $S_\mathrm{m}$ around $T_{\mathrm C}$ calculated from specific heat data in $0\,\mathrm{T}$ (black curve) and $1\,\mathrm{T}$ with field applied along (blue curve) and perpendicular to the $c$-axis (red curve). (b) The magnetic entropy change $\Delta S_\mathrm{m}=S_\mathrm{m}(H)-S_\mathrm{m}(0)$ obtained from specific heat (solid curves) and magnetic measurements (symbols), with field applied along (blue) and perpendicular to the $c$-axis (red).}
	\label{Fe2P_Entropy}
\end{figure}

The effects of the applied magnetic fields are well illustrated through the magnetic entropy. The magnetic entropy $S_\mathrm{m}$ can be obtained from the specific heat measurements of Fig.~\ref{Fe2P_Calori} as the integral of $C_\mathrm{m}/T$ over temperature. Putting $S_\mathrm{m}(300\,\mathrm{K})=0$ as a reference point, the thus obtained curves for $H=0$ and $H=1\,\mathrm{T}$ applied along and perpendicular to the $c$-axis are shown in Fig.~\ref{Fe2P_Entropy}(a).
In zero field (black curve), the entropy increase $\Delta S_\mathrm{L}$ due to the latent heat is added as a step function at $T_\mathrm{C}$. It is seen that the magnetic entropy for $\mathbf{H}\parallel\,\bf{c}$  (blue curve) is lower than that of zero field, both above and below $T_\mathrm{C}$. For $\mathbf{H}\perp\,\bf{c}$ (hard axis) the entropy is higher than for $\mathbf{H}\parallel\,\bf{c}$ at all temperatures.

Interestingly, $S_\mathrm{m}(1\,\mathrm{T})$ is crossing $S_\mathrm{m}(0\,\mathrm{T})$ at $T_\mathrm{C}$, making the low-temperature entropy for $\mathbf{H}\perp\,\bf{c}$ higher than that of zero field (Fig.~\ref{Fe2P_Entropy}(a)). This effect is clearly seen in Fig.~\ref{Fe2P_Entropy}(b), showing the magnetic entropy change $\Delta S_\mathrm{m}=S_\mathrm{m}(1\,\mathrm{T})-S_\mathrm{m}(0\,\mathrm{T})$. The magnetic entropy change $\Delta S_\mathrm{m}$ due to a magnetization process can be calculated from magnetization data (Fig.~\ref{Fe2P_Magnetization}) using Maxwell relations ($\Delta S_\mathrm{m}(T,H)=\mu_0\int_0^{H} ({dM(T,H)}/{dT})\cdot dH$) \cite{Tishin03}. The thus derived $\Delta S_\mathrm{m}$ are shown as open symbols in Fig.~\ref{Fe2P_Entropy}(b). 
The general behaviors of $\Delta S_\mathrm{m}$ obtained from specific heat and magnetization are in good agreement, with the results from magnetic measurements giving a value $\sim$10\% higher at the position of the zero field transition. 

For $\mathbf{H}\parallel\,\bf{c}$, a strong magnetocaloric effect (entropy is decreasing with increasing field) is seen as a dip in $\Delta S_\mathrm{m}$. However, for $\bf{H}\, \perp \bf{c}$, a positive $\Delta S_\mathrm{m}$ is found below $T_{\mathrm C}$, corresponding to an \emph{inverse} magnetocaloric effect. This positive magnetic entropy change reflects an increased disorder of the spin system in the temperature range where the applied magnetic field and the effective magneto-crystalline anisotropy are of equal strength (near the cusps in $M$ vs. $T$). Increased disorder of the spin system should be reflected in a decreased total magnetization with increasing field in this area. This is indeed seen from measurements of the total magnetization \cite{Caron13b}.

\begin{figure}[b!]
	\includegraphics[width=0.594\linewidth]{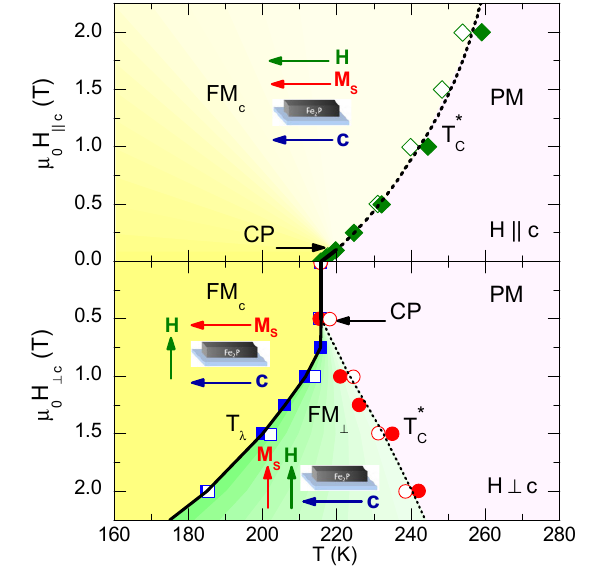}
	\caption{Magnetic phase diagram for Fe$_2$P obtained from specific heat (filled symbols) and magnetization measurements (open symbols), complemented with published results~\cite{Caron13b}.
The upper panel shows the phase diagram for magnetic field applied along the $c$-axis, while the lower panel shows the case for $\bf{H}\,\perp\,\bf{c}$ (with reversed ordinate). The direction of $M_{\mathrm s}$, is indicated for each ferromagnetic phase. The transition is first order for fields below the directional-dependent critical point, marked CP.}
	\label{Fe2P_PhaseDiag}
\end{figure}

The derived $H-T$ magnetic phase diagram for Fe$_2$P is shown in Fig.~\ref{Fe2P_PhaseDiag}. At low fields $\bf{H}\,\parallel\,\bf{c}$ (upper panel), $T_{\mathrm C}$ is shifted to higher temperatures, forming a line of first order phase transitions that ends at a critical point $\mu_0 H_\mathrm{\parallel,cr} < 0.1\,\mathrm{T}$. Above the critical point, the continuation of the phase transition is determined from the maximum in specific heat at different applied fields (filled symbols in Fig.~\ref{Fe2P_PhaseDiag}) and the maximum of $dM/dT$ at constant field (open symbols), agreeing well with each other. For the case of $\bf{H}\,\perp\,\bf{c}$, lower panel of Fig.~\ref{Fe2P_PhaseDiag}, the first order transition extends up to a critical or possibly tricritical point $\mu_0 H_\mathrm{\perp,cr} \approx 0.5\,\mathrm{T}$. For fields above $H_\mathrm{\perp,cr}$ two characteristic temperatures are observed in the magnetization and specific heat, indicated by $T^\star_{\mathrm C}$ (increasing with increasing magnetic field) and $T_{\lambda}$ (decreasing with increasing field) in Fig.~\ref{Fe2P_PhaseDiag}. The minimum in $dM/dT$ closely coincides with $T^\star_{\mathrm C}$ as obtained from specific heat. The lower characteristic temperature corresponding to $T_{\lambda}$ is taken as the maximum in $dM/dT$ below the peak in magnetization (see Fig.~\ref{Fe2P_Magnetization}), which decreases with increasing field and roughly coincides with the $T_{\lambda}$ found from specific heat, as seen in the lower panel of Fig.~\ref{Fe2P_PhaseDiag}.

The magnetic entropy change $\Delta S_\mathrm{L}$ at the first-order part of the transition line $T^\star_{\mathrm C}(H)$ is related to the jump in magnetization $\Delta M$ and slope of the phase boundary through the magnetic analogue of the Clausius-Clapeyron equation: ${\mathrm{d}T^\star_{\mathrm C}}/{\mathrm{d}(\mu_0H)} =-{\Delta M}/{\Delta S_\mathrm{L}}$. With $\Delta M$ from Fig.~\ref{Fe2P_Magnetization} and $\Delta S_\mathrm{L}$ from the latent heat measurements, we calculate the initial slope of the phase transition line to be $59\,\mathrm{K/T}$. From the data shown in Fig.~\ref{Fe2P_PhaseDiag} we find $\mathrm{d}T^\star_{\mathrm C}/\mathrm{d}(\mu_0 H) = 38\,\mathrm{K/T}$ for fields $\leq 0.1\,{\rm T}$ along the $c$-axis.
For $\bf{H}\,\perp\,\bf{c}$, on the other hand, there is no magnetization jump along the applied field at $T^\star_{\mathrm C}$. Thus, $T^\star_{\mathrm C}$ is independent of $H$ for this field direction, up to the critical point $\mu_0 H_\mathrm{\perp,cr}$ where $\Delta S_\mathrm{L}$ goes to zero.

To interpret the full phase diagram, first note that the zero-field transition is a combined elastic and magnetic transition from a paramagnetic (PM) phase into a ferromagnetic phase ordered along the $c$-axis (FM$_\mathrm{c}$). For $\bf{H}\,\parallel\,\bf{c}$ (Fig.~\ref{Fe2P_PhaseDiag}), it is clear that magnetic fields applied in this direction stabilize the FM$_\mathrm{c}$ phase. For $\bf{H}\,\perp\,\bf{c}$ at low fields ($H<H_\mathrm{\perp,cr}$), the location of the PM-to-FM$_\mathrm{c}$ phase boundary is unaffected by the applied field, similarly indicating that its spontaneous magnetization direction is perpendicular to the field in this case. However, at higher fields a new phase appears between $T_{\lambda}$ and $T^\star_{\mathrm C}$, indicated as FM$_\perp$. This phase represents a state \emph{without} spontaneous magnetization $M_\mathrm{s}$ along the $c$-axis, but with ferromagnetic order stabilized by the applied field. At the transition line $T_{\lambda}$ the nature of the anisotropy changes from a pure uniaxial character with linear increase of the magnetization with increasing perpendicular fields into a regime with higher order anisotropy terms. This can be seen from the $M$ vs. $H$ curves in Fig.~\ref{Fe2P_Magnetization}(a) where an increase of the field causes further alignment of the magnetization along the applied field. Such a behavior is different from the reordering transition discussed in literature~\cite{Reis08, Becerra2000}, where the spin reordering has a first-order nature. The elastic transition is closely following the onset of magnetic ordering along the $c$-axis, i.e.,  along $T^\star_{\mathrm C}$ for $\bf{H}\,\parallel\,\bf{c}$ and $T_\lambda$ for $\bf{H}\,\perp\,\bf{c}$. Since the elastic transition does not bring much latent heat, it is likely that it is driven by the FM$_\mathrm{c}$ ordering.

In conclusion, we have found that the low-field FOMT of Fe$_2$P ends at a critical point governed by strong magneto-crystalline anisotropy. For fields applied perpendicular to the $c$-axis, the low-temperature ferromagnetic state displays increasing magnetic entropy with increasing field, corresponding to a negative magnetocaloric effect. At higher temperatures and fields a new phase FM$_\perp$ appears, where the $c$-axis anisotropy becomes ineffective. This implies an orientational order (along the $c$-axis) to disorder transition at the $T_{\lambda}$ phase line. The structural (elastic) transition is found to be linked to this $c$-axis ordering and exhibits associated latent heat at low fields $\lesssim10\%$ of the total latent heat.

\begin{acknowledgments}
Financial support from the Swedish Research Council is acknowledged (M.\,H., contract 2012-6562). We thank M.\,Fiebig (ETH) and O.\,Tjernberg (KTH) for hosting M.\,H. during his postdoc, and E.\,Br\"{u}ck, Z.\,Diao and V.\,M.\,Krasnov for useful discussions.
\end{acknowledgments}


\end{document}